\documentstyle[graphicx,amsmath,amssymb,fleqn]{mn}

\newif\ifAMStwofonts
\AMStwofontstrue
\DeclareMathAlphabet{\mathfrak}{U}{euf}{m}{n}
\SetMathAlphabet\mathfrak{bold}{U}{euf}{b}{n}
\newcommand{\bfr}{{\bmath r}}
\newcommand{\bfv}{{\bmath v}}

\newcommand{\kelvin}{{\mbox{\,K}}}
\newcommand{\micron}{{\mbox{\,$\mu$m}}}
\makeatletter
  \newcommand\figcaption{\def\@captype{figure}\caption}
  \newcommand\tabcaption{\def\@captype{table}\caption}
\makeatother
\renewcommand{\leq}{\leqslant}

\title[Kinematics of dusty ellipticals]%
{Kinematics of elliptical galaxies with a diffuse dust component}

\author[Baes \& Dejonghe]%
{Maarten Baes$^1$\thanks{Research Assistent of the Fund for Scientific Research - Flanders (Belgium)} and Herwig Dejonghe$^1$\\
$^1$Sterrenkundig Observatorium Universiteit Gent, Krijgslaan 281 S9, B-9000 Gent, Belgium
} 

\begin{document}

\maketitle
\begin{abstract}
Observations show that early-type galaxies contain a considerable
amount of interstellar dust, most of which is believed to exist as a
diffusely distributed component. We construct a four-parameter
elliptical galaxy model in order to investigate the effects of such a
smooth absorbing component on the projection of kinematic quantities,
such as the line profiles and their moments. We investigate the
dependence on the optical depth and on the dust geometry. Our
calculations show that both the amplitude and the morphology of these
quantities can be significantly affected. Dust effects should
therefore be taken in consideration when interpreting photometric and
kinematic properties, and correlations that utilize these quantities.
\end{abstract}

\section{Introduction}

Theoretical work on the transfer of radiation in the interstellar
medium demonstrated that the effects of dust grains are dominant over
those of any other component (Mathis 1970; Witt {\em et al.\ }1992,
hereafter WTC92). From the observational point of view, dust in
elliptical galaxies is optically detectable by its obscuration effects
on the light distribution, when it is present in the form of dust
lanes and patches (e.g.\ Bertola \& Galetta 1978, Hawarden {\em et
al.\ }1981, Ebneter \& Balick 1985, Sadler \& Gerhard 1985). In
emission, cold dust ($T_d\approx30\kelvin$) is detected at
far-infrared wavelengths. The {\em IRAS} satellite detected dust in
ellipticals in the 60 and 100\micron\ wavebands (e.g.\ Jura 1986,
Bally \& Thronson 1989, Knapp {\em et al.\ }1989) in, at that time,
unexpectedly large quantities.  However, {\em IRAS} is insensitive to
very cold dust ($T_d<30\kelvin$), and preliminary {\em ISO} results
(Haas 1998) and (sub)millimeter data (Fich \& Hodge 1993, Wiklind \&
Henkel 1995) may indicate the presence of an additional very cold dust
component in ellipticals.

All these observations show that the presence of dust in ellipticals
is the rule rather than the exception (van Dokkum \& Franx 1995,
Roberts {\em et al. }1991). A straightforward question is whether the
dust detected in emission and absorption represents one single
component. Goudfrooij \& de Jong (1995) show that the {\em IRAS} dust
mass estimates are roughly an order of magnitude higher than those
calculated from optical data. To solve this discrepancy, they
postulated a two-component interstellar dust medium. The smaller
component is an optically visible component in the form of dust lanes,
the larger one is diffusely distributed over the galaxy, and therefore
hard to detect optically. At least part of this smooth dust
distribution can be easily explained in ellipticals : the old stellar
populations have a significant mass loss rate of gas and dust (Knapp,
Gunn \& Wynn-Williams 1992), which one expects to be distributed
smoothly over the galaxy.

The effects of a diffuse dust component on the photometry are studied
by WTC92 and Wise \& Silva (1996, hereafter WS96). They demonstrate
that even modest amounts of dust can produce significant broadband
colour gradients, without changing the global colours. Radial colour
gradients can therefore indicate the presence of this diffuse dust
component.

Hitherto the effect of a diffuse absorbing medium on the internal
kinematics of ellipticals has been neglected, probably due to the
widespread idea that early-type galaxies are basically dust-free. In
this paper, we construct a set of elliptical galaxy models in order to
investigate these effects. Our models contain 4 parameters, allowing
us to cover a wide range in orbital structure, optical depth and
relative geometry of the dust and the stellar component. 

We want to demonstrate the qualitative and quantitative effects of
dust absorption on the observed kinematic quantities. In the stellar
dynamical view of a galaxy, the ultimate purpose is the determination
of the phase space distribution function $F(\bfr,\bfv)$, describing
the entire kinematic structure. This distribution function is usually
constructed by fits to observed kinematic profiles as the projected
density $\rho_p(x)$ and the projected velocity dispersion
$\sigma_p^2(x)$. Using a completely analytical spherical model for
ellipticals, we demonstrate in this paper the effects of dust on these
projected profiles and on line profiles.
 
In section~2 we describe the mechanism of radiative transfer and the
effect of dust on the projection of physical quantities. In section~3
we describe our model in detail. The results of our calculations are
presented and discussed in section~4. Section~5 sums up.

\section{Dusty projections}

A fundamental problem in the study of stellar systems is to retrieve
three-dimensional information from a two-dimensional image. This
deprojection problem is degenerate and can only be solved with
specific assumptions, e.g.\ spherical symmetry. In this section we
show how dust absorption affects the projection. In this first study
we limit ourselves to absorption effects and we do not include
scattering.

The transfer equation describes how radiation and matter interact. The
radiation field at a point $\bfr$ propagating in a direction
${\bmath n}$ is characterized by the specific intensity
$I(\bfr,{\bmath n})$, while the global optical properties of the
galaxy are described by the light density $\rho(\bfr)$ and the opacity
$\kappa(\bfr)$. The transfer equation on a line-of-sight $p$ is then
\begin{equation}
	\frac{{\rm d}I}{{\rm d}s}(p;s) 
	= 
	\rho(\bfr)-\kappa(\bfr)\,I(p;s),
\label{EOT}
\end{equation}
where $s$ is the path length on $p$ and $I(p;s)\equiv I(\bfr,{\bmath
e}_s)$ is the intensity at the point $s$ on $p$ propagating in the
direction of the observer ${\bmath e}_s$. Equation~(\ref{EOT}) can be
readily solved to obtain the light profile detected by the observer
$\rho_p(p)\equiv I(p;s_o)$,
\begin{equation}
	\rho_p(p) = \int^{s_o}\!\rho(\bfr)\,
	\exp\left(-\int_{s}^{s_o}\!\kappa(\bfr)\,{\rm d}s'\right) 
	{\rm d}s,
\label{generalrho}
\end{equation}
where $s_o$ denotes the location of the observer and the integration
covers the entire path. Equation~(\ref{generalrho}) can easily be
generalized to include kinematic information:
\begin{equation}
	f_p(p,\bfv) = \int^{s_o}\!f(\bfr,\bfv)\,
	\exp\left(-\int_{s}^{s_o}\!\kappa(\bfr)\,{\rm d}s'\right) 
	{\rm d}s.
\label{generalf}
\end{equation}
This expression yields what we call the dusty projection $f_p(p,\bfv)$
of the kinematic quantity $f(\bfr,\bfv)$. 

\begin{figure}
\centering \includegraphics[clip,bb=28 230 568 610,width=85mm]%
{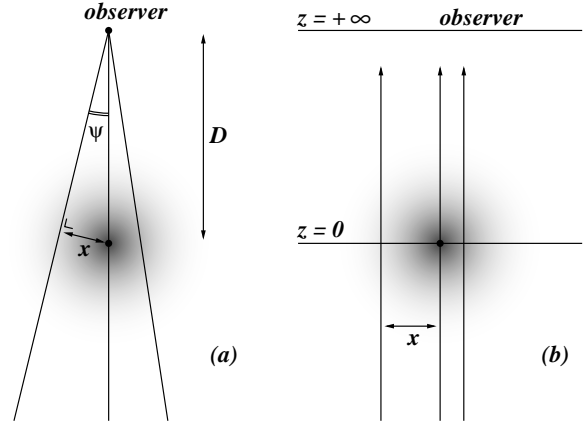}
\caption{The two coordinate systems as described in section~2. When
the distance $D$ between the observer and the galaxy's centre goes to
infinity, figure~{(a)} transforms into figure~{(b)}.}
\label{coordsys2.ps}
\end{figure}

To describe the observed kinematics of spherical galaxies we adopt a
coordinate system as illustrated in figure~{\ref{coordsys2.ps}{a}}. The
distance between the observer and the centre of the system is denoted
by $D$. Due to the spherical symmetry we can write
$f(\bfr,\bfv)=f(r,\bfv)$ and $\kappa(\bfr)=\kappa(r)$, with
$r=\Vert\bfr\Vert$ the spherical radius. Lines-of-sight are determined
by the angle $\psi$ they make with the line-of-sight through the
system's centre. Substituting this in equation~(\ref{generalf}), one
finds
\begin{multline}
	f_p(\psi,\bfv)
	=
	\exp\left(
	-\int_x^D\frac{\kappa(r)\,r\,{\rm d}r}
	{\sqrt{r^2-x^2}}\right) \times
	\\
	\qquad\quad\left[\int_x^D\exp\left(
	\int_x^r\frac{\kappa(r')\,r'\,{\rm d}r'}{\sqrt{{r'}^2-x^2}}\right)
	\frac{f(r,\bfv)\,r\,{\rm d}r}{\sqrt{r^2-x^2}}
	\right.
	\\ 
	+\left.\int_x^{+\infty}\!\!\exp\left(
	-\int_x^r\frac{\kappa(r')\,r'\,{\rm d}r'}{\sqrt{{r'}^2-x^2}}\right)
	\frac{f(r,\bfv)\,r\,{\rm d}r}{\sqrt{r^2-x^2}}
	\right]
	\label{centrproj}
\end{multline}
where we have set $x=D\sin\psi$. 

This equation can be simplified substantially by making the
approximation of parallel projection, as in
figure~{\ref{coordsys2.ps}{b}}. Since the scalelengths of galaxies is
several orders of magnitude smaller than their distance, the induced
error of the order $(x/D)^2$ is usually negligible. In this system a
line-of-sight $p$ is determined by its projected radius $x$ on the
plane of the sky. Equation~(\ref{generalf}) becomes
\begin{equation}
	f_p(x,\bfv) 
	=
	\int_{-\infty}^{+\infty}\!\!f(r,\bfv)\,
 	\exp\left(-\int_z^{+\infty}\!\!\kappa(r)\,{\rm d}z'\right)
	{\rm d}z.
\label{lightprofile}
\end{equation}
We can rewrite the integrals in~(\ref{lightprofile}) as integrals over
$r$ and obtain
\begin{subequations}
\begin{equation}
	f_p(x,\bfv)
	=
	2\int_x^{+\infty}\!{\cal K}(x,r)\,
	\frac{f(r,\bfv)\,r\,{\rm d}r}{\sqrt{r^2-x^2}},
\label{dusty_projection}
\end{equation}
where
\begin{gather}
	{\cal K}(x,r) 
	=
	\exp\left(-\int_x^{+\infty}
	\frac{\kappa(r)\,r\,{\rm d}r}{\sqrt{r^2-x^2}}\right) 
	\nonumber \\
	\qquad\qquad\qquad\qquad \times \quad
	\cosh\left(\int_x^r
	\frac{\kappa(r')\,r'\,{\rm d}r'}{\sqrt{{r'}^2-x^2}}\right).
\label{weight_function}
\end{gather}
\end{subequations}
The same result can simply be found by taking the limit
$D\rightarrow+\infty$ in expression~(\ref{centrproj}).

The weight function ${\cal K}(x,r)$ assumes values between 0 and 1,
and reduces to unity if the opacity vanishes, in correspondence with
the normal, dust-free projections. In the sequel of this paper we will
always refer to parallel projection.

\section{The model}

\subsection{The stellar component}

Since dusty projections already require a double integration along the
line-of-sight, one is inclined to use a dynamical model that is as
simple as possible to avoid cumbersome numerical work. Therefore the
Plummer model (Dejonghe 1987), representing a spherical cluster
without central singularity, seems an obvious candidate. Although the
photometry of this model does not fit real elliptical
galaxies\footnote[1]{A more thorough analysis of the effects of
diffuse dust on more realistic photometric profiles can be found in
WTC92 and WS96.}, it has the huge advantage that the entire kinematic
structure is sufficiently rich to include radial and tangential
models, and that it can be described analytically. We can safely
assume that our results will be generic for the class of elliptical
galaxies.

We fix the units by choosing a value for the total stellar mass $M$,
the gravitational constant $G$ and a scale length $r_*$. Working in
dimensionless units we choose $M=G=r_*=1$. For the potential we choose
the natural unit $\psi_* = G\,M/r_*$ and we set $\psi(0)=1$. In these
units the central escape velocity equals $\sqrt{2}$, and the Plummer
model is defined by
\begin{subequations}
\begin{gather}
	\psi(r) = \frac{1}{\sqrt{1+r^2}}
	\label{plummpair1} \\
	\rho(r) = \frac{3}{4\pi}\,
		  \left(1+r^2\right)^{-\frac{5}{2}}.
	\label{plummpair2}
\end{gather}
\end{subequations}
Dejonghe (1986) describes a technique to construct 2-integral
distribution functions $F(E,L)$ that self-consistently generate a
given potential-density pair, based on the construction of an
augmented mass density $\tilde{\rho}(\psi,r)$, i.e.\ the density
expressed as a function of $r$ and $\psi$. For the potential-density
pair~(\ref{plummpair1},b) the following one-parameter family is
convenient
\begin{equation}
	\tilde{\rho}(\psi,r) 
	= 
	\frac{3}{4\pi}\,\psi^{5-q}\,\left(1+r^2\right)^{-\frac{q}{2}}.
\label{rhotilde}
\end{equation}
This results in a family of completely analytical anisotropic Plummer
models, fully described by Dejonghe (1987). The distribution function
$F_q(E,L)$ and its moments can be expressed analytically. For example,
the radial and tangential velocity dispersions read
\begin{subequations}
\begin{gather}
	\sigma_r^2(r) 
	= 
	\frac{1}{6-q}\frac{1}{\sqrt{1+r^2}}
	\label{raddisp} \\
	\sigma_{\phi}^2(r) 
	=
	\sigma_{\theta}^2(r)
  	=
	\frac{1}{6-q}\frac{1}{\sqrt{1+r^2}}    
	\left(1-\frac{q}{2}\,\frac{r^2}{1+r^2}\right),
\end{gather}
\end{subequations}
and Binney's anisotropy parameter equals
\begin{equation}
	\beta(r)
	=
	1 - \frac{\sigma_{\phi}^2}{\sigma_r^2}
	=
	\frac{q}{2}\,\frac{r^2}{1+r^2}.
\end{equation}
This expression provides the physical meaning of the parameter $q$. It
has the same sign as $\beta(r)$, and its range is restricted to
$-\infty<q\leq2$ (since $-\infty<\beta(r)\leq1$). Tangential clusters
have $q<0$, radial clusters have $0<q\leq2$ and for $q=0$ the model is
isotropic. For small radii all models are fairly isotropic, while the
true nature of the orbital structure becomes clearly visible in the
kinematics at radii $r\gg1$. More specifically, we consider a
tangential model ($q=-6$), a radial one ($q=1$) and the isotropic case
($q=0$).

\subsection{The dust component}

A dust model is completely determined by its opacity function
$\kappa(r)$. Besides the geometrical dependency, this function is also
dependent upon the wavelength, which we do not mention
explicitly. WTC92 considered an elliptical galaxy model with a
diffusely distributed dust component. They used a King profile for
both the stellar and the dust component. Analogous models were
considered by Goudfrooij \& de Jong (1995) and WS96. In the assumption
that dust is only present in the inner part of ellipticals, WTC92 cut
off the dust distribution at two thirds of the galaxy
radius. Nevertheless dust may also be present in the outer regions,
with temperatures of the order of 20\,K (Jura 1982, Goudfrooij \& de
Jong 1995).

Our dust model is defined by the opacity function 
\begin{equation}
	\kappa(r)
	=
	\frac{1}{\sqrt{\pi}}\,\tau\,
	\frac{\Gamma\left(\frac{\alpha}{2}\right)}
	{\Gamma\left(\frac{\alpha-1}{2}\right)}\,
	\frac{1}{r_0}\,
	\left(1+\frac{r^2}{r_0^2}\right)^{-\frac{\alpha}{2}},
\label{king_para}
\end{equation}
with $r_0$ a scale factor and $\alpha$ the so-called dust
exponent. The normalization is such that $\tau$ equals the total
optical depth (in the $V$-band), defined as
\begin{equation}
	\tau = 2\int_0^{+\infty}\kappa(r)\,{\rm d}r.
\label{normalization}
\end{equation}
We can calculate the weight function ${\cal K}(x,r)$ by
substituting~(\ref{king_para}) in~(\ref{weight_function}),
\begin{equation}
	{\cal K}(x,r) 
	=
	\exp\left(-\frac{\tau(x)}{2}\right)\,
	\cosh\left[
	\frac{\tau(x)}{2}\,{\bf U}_{\alpha-3}\!\left(
	\sqrt{\frac{r^2-x^2}{r^2+r_0^2}}\right)
	\right],
\label{kingweight}
\end{equation}
where $\tau(x)$ is the optical depth along a line-of-sight $x$
\begin{equation}
	\tau(x) 
	= 
	2\,\int_x^{+\infty}\frac{\kappa(r)\,r\,{\rm d}r}
	{\sqrt{r^2-x^2}}
	=
	\tau\,\left(1+\frac{x^2}{r_0^2}\right)^{-\frac{\alpha-1}{2}},
\label{optdepth}
\end{equation}
and the function ${\bf U}_\alpha(z)$ is defined in
appendix~{\ref{appendixA}}.

\subsection{Parameter space}

\subsubsection{The optical depth}

The optical depth is an indicator of the amount of dust in the galaxy,
and is therefore the primary parameter in the models. WS96 estimated
the optical depths of a sample of ellipticals by fitting radial colour
gradients and found a median value of $\tau\la2$\footnote[1]{This is
scaled to the definition of $\tau$ adopted in this paper, defined as
the projection of the opacity along the central line-of-sight through
the entire galaxy. WS96 defined $\tau$ as the integral of the opacity
from the centre of the galaxy to the edge, half our value.}. We
explore optical depths between $\tau=0.1$ and $\tau=200$.

\subsubsection{The dust geometry}

The parameters $r_0$ and $\alpha$ determine the shape of the dust
distribution. The exponent $\alpha$ is restricted to
$\alpha>1$. Smaller values of $\alpha$ or larger values of $r_0$
correspond to extended distributions; for larger values of $\alpha$ or
smaller values of $r_0$ the dust is more concentrated in the central
regions. Since the two geometry parameters $r_0$ and $\alpha$ have
roughly the same effect, we concentrate upon the case $r_0=1$ and we
take $\alpha$ as the parameter determining the geometry.

For high values of $\alpha$ the distribution is very concentrated in
the central regions of the galaxy. {\em HST} imagery has revealed
compact dust features in the majority of the nearby ellipticals (van
Dokkum \& Franx 1995), and these could be an indication of an
underlying concentrated diffuse dust distribution. However, Silva \&
Wise (1996) showed that such smooth dust concentrations should induce
easily detectable colour gradients in the core, even for small optical
depths. But high resolution {\em HST} imaging of two samples of nearby
ellipticals revealed that elliptical galaxy cores have usually small
colour gradients (Crane {\em et al.\ }1993, Carollo {\em et al.\
}1997), and thus no direct indication of concentrated diffuse dust is
found. Therefore high values of $\alpha$ seem to be less probable and
we will only treat them for the sake of completeness.

On the other hand, for small values of the dust component the
distribution will be extended. If $\alpha$ approaches 1 the
distribution is very extended, and little dust is located in the
central regions of the galaxy. In the limit the dust is infinitely
thin distributed over all space, the contribution of the dust in the
galaxy is zero and the dust effectively forms an obscuring layer
between the galaxy and the observer. This geometry is known as the
overlying screen model (Holmberg 1975, Disney {\em et al.\ }1989) and
is the analogue for extinction of galaxian stars. By taking the limit
$\alpha\rightarrow1$ in~(\ref{optdepth}) we obtain
\begin{equation}
	\tau(x) = \tau
\end{equation}
and substitution in~(\ref{kingweight}) yields
\begin{equation}
	{\cal K}(x,r) = \exp\left(-\frac{\tau}{2}\right).
\label{screenweight}
\end{equation}
Notice that the same result can be found by taking the limit
$r_0\rightarrow+\infty$.

As demonstrated in section~2, the computational prize of dust
absorption on the projection of kinematic quantities is a double
integration along the line-of-sight instead of a single one. This can
be avoided if the weight function ${\cal K}(x,r)$ can be calculated
analytically for all $x$ and $r$, i.e.\ if the functions ${\bf
U}_{\alpha-3}(z)$ can be evaluated explicitly. For obvious
computational reasons, we concentrate on the cases where $\alpha$ is
an integer number. But to study the effects in the neighborhood of the
singular value $\alpha=1$, we also consider a range of non-integer
values for $\alpha$ between 1 and 2.

\section{The projected kinematics}

For the dust-free model all the moments of a line profile can be
calculated analytically (see appendix~{\ref{appendixC}}). For the
lower order moments we find
\begin{equation}
	\rho_p(x) 
	= 
	\frac{1}{\pi}\frac{1}{\left(1+x^2\right)^2},  
\label{plummer:projdens}
\end{equation}
and
\begin{equation}
	\sigma_p^2(x) 
	= 
	\frac{9\pi}{32}\,\frac{1}{6-q}\,
	\frac{1}{\sqrt{1+x^2}}
	\left(1-\frac{5\,q}{12}\,\frac{x^2}{1+x^2}\right).
\label{plummer:projdisp}
\end{equation}
For dusty models this projection has to be carried out numerically. We
calculated the projected mass density, the projected velocity
dispersion and a set of line profiles for our dusty galaxy models. In
this section we compare the normal and the dusty projections of the
kinematic quantities, and we investigate both the dependence upon the
optical depth $\tau$ and upon the dust geometry.

\subsection{Dependence upon optical depth}
\label{deptau.par}

In this section we keep the dust geometry fixed and vary the optical
depth $\tau$. We adopt a modified Hubble profile ($\alpha=3$), which
is slightly more extended than the stellar light.

\subsubsection{The projected density}

\begin{figure*}
\centering 
\includegraphics[clip,bb=60 350 536 685]{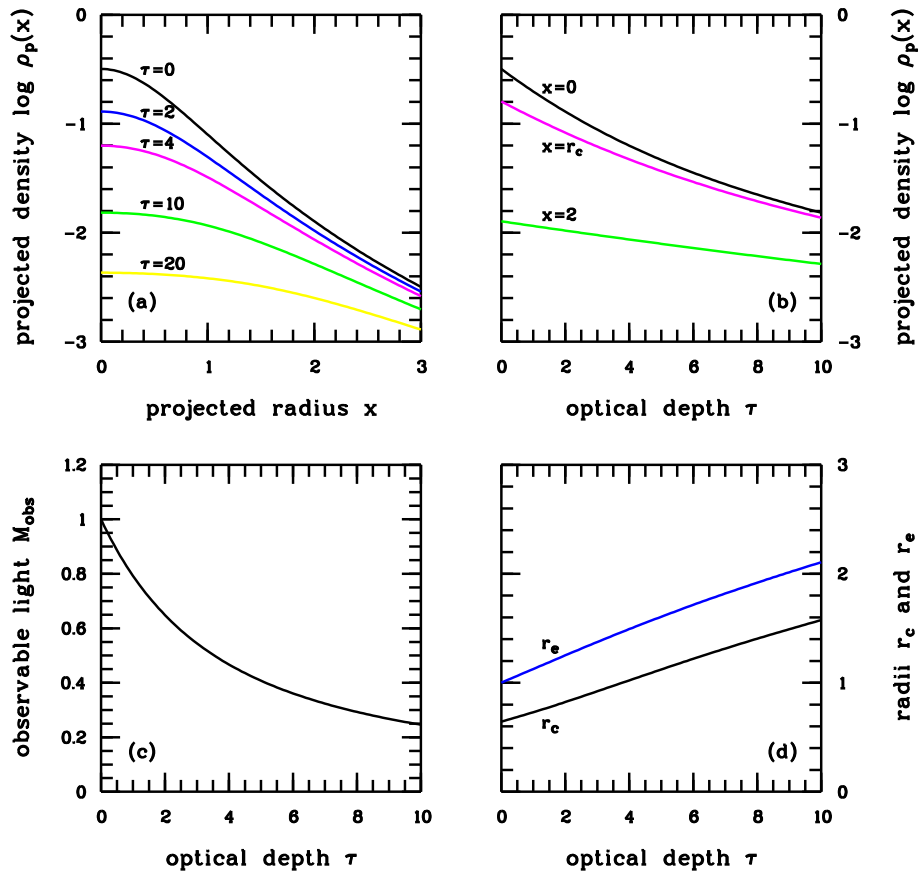}
\caption{The dependence of the projected density $\rho_p(x)$ upon the
optical depth $\tau$. The dust model is the $\alpha=3$ model. In
panel~{(a)} the projected density is plotted as a function of the
projected radius $x$ for various values of $\tau$, which are
indicated. In figure~{(b)} we show the projected density explicitly as
a function of the optical depth $\tau$, for three lines-of-sight
$x$. The curve for $x=0$ can be calculated explicitly, as done in
appendix~{\ref{appendixC}}. The lower two panels show {(c)} the total
integrated light $M_{\rm obs}$, and {(d)} the core radius $r_c$ and the
effective radius $r_e$ as a function of $\tau$.}
\label{deptaurho.ps}
\end{figure*}

The primary effect of dust absorption on the projected density profile
(figure~{\ref{deptaurho.ps}{a,b}) is obvious~: an overall
reduction. Even for modest amounts of dust this effect is clearly
observable. The projected density in the central regions is already
reduced to half of its original value for an optical depth
$\tau=1.5$. At the core radius $r_c=0.64$ this occurs for
$\tau=3$. The global effect of the extinction is shown in
figure~{\ref{deptaurho.ps}{c}, which displays the observed
integrated light $M_{\rm obs}$, i.e.\ the fraction of the total light
output that can be detected by the observer. It is obtained by
integrating the projected density profile $\rho_p(x)$ over the entire
plane of the sky,
\begin{equation}
	M_{\rm obs} = 2\pi\,\int_0^{+\infty}\rho_p(x)\,x\,{\rm d}x.
\end{equation}
The global extinction effect is strong~: for an optical depth $\tau=1$
already 20 per cent of the mass is hidden for the observer, and for
$\tau=3.4$ this becomes 50 per cent.

Besides reducing the amount of starlight, the dust has also a
qualitative effect on the projected density profile. Since the dust is
primarily located in the central portions of the galaxy, the
$\rho_p(x)$ profile will be flattened towards the centre. As shown by
WTC92 and WS96 this causes the production of colour gradients. But it
has also a non-negligible effect on all global photometric quantities,
e.g.\ luminosities and core and global diameters. For example, the
apparent core of the galaxy (quantified by the core radius $r_c$ or
the effective radius $r_e$) will be larger due to the central
flattening (figure~{\ref{deptaurho.ps}{d}}). This effect can be up to
25 per cent or more for our basic model with moderate optical depths
$\tau\approx2$. On the other hand the total diameter of the galaxy
(usually expressed as the diameter of the 25\,mag\,arcsec$^{-2}$
isophote) will appear smaller due to the global attenuation.

\subsubsection{The projected dispersion}

\begin{figure*}
\centering 
\includegraphics[clip,bb=60 350 536 685]{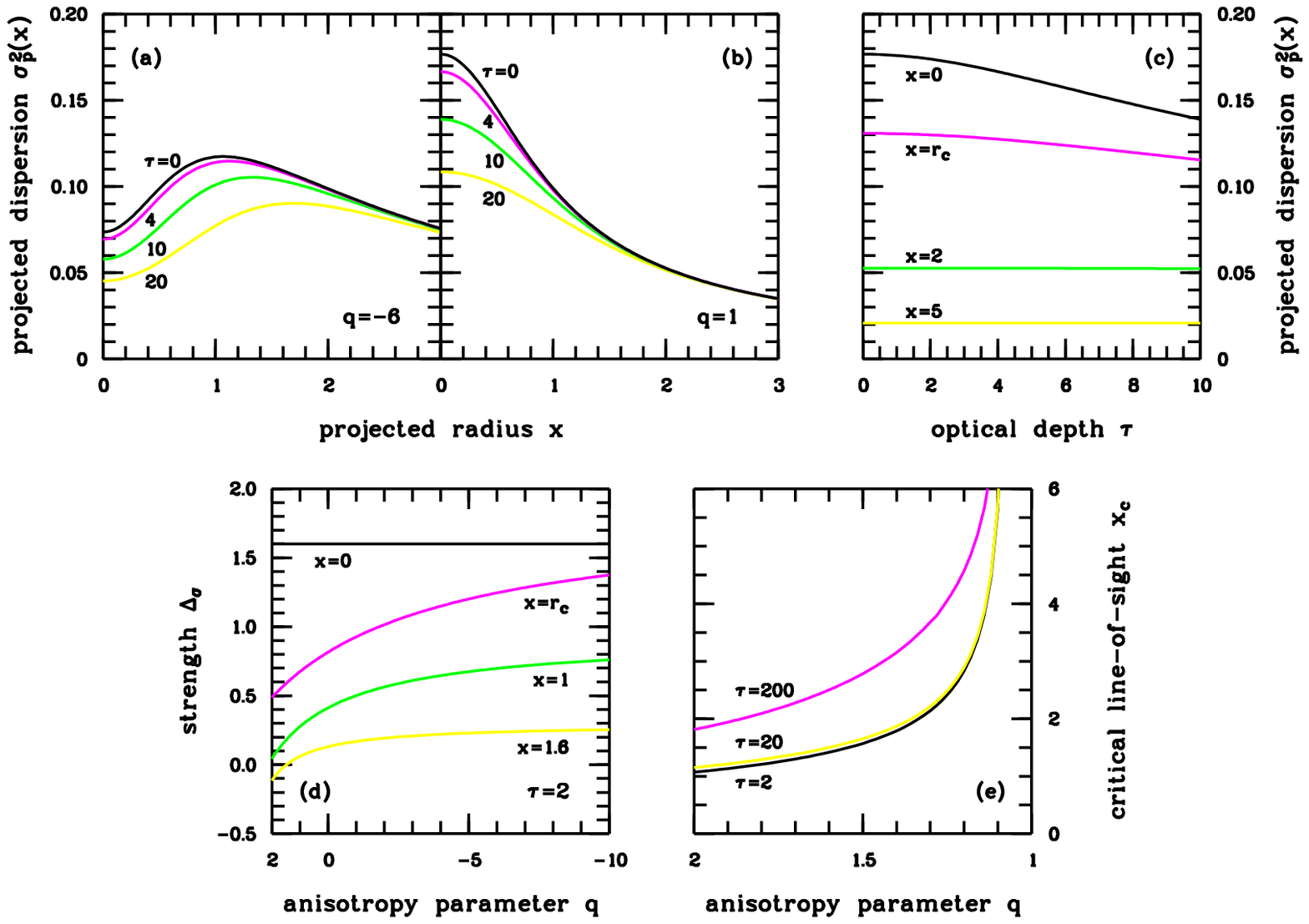}
\caption{The dependence of the projected dispersion $\sigma_p^2(x)$
upon the optical depth $\tau$. The dust model is the $\alpha=3$
model. In the upper two panels the projected dispersion is plotted as
a function of the projected radius $x$ for various values of $\tau$,
which are indicated. It is shown {(a)} for a tangential model with
$q=-6$ and {(b)} a radial model with $q=1$. In panel~{(c)} we
show the projected dispersion explicitly as a function of the optical
depth $\tau$, for four lines-of-sight $x$. It is plotted for the
radial model $q=1$. The curve for $x=0$ can be calculated explicitly,
as done in appendix~{\ref{appendixC}}. In figure~{(d)} we plot the
strength $\Delta_\sigma(x)$ of the effect on the dispersion profile as
a function of the anisotropy parameter $q$ for three different
lines-of-sight, which are indicated. The optical depth is
$\tau=2$. The last plot~{(e)} displays the critical line-of-sight
$x_c$ for which $\Delta_\sigma(x_c)=0$ as a function of the anisotropy
parameter $q$, for different values of $\tau$. Above these curves the
parameter has negative values, below are positive values.}
\label{deptausig.ps}
\end{figure*}

The effect of dust on the projected dispersion profile $\sigma_p^2(x)$
has a totally different nature than the effect on the projected
density profile $\rho_p(x)$, since it is a normalized projection. For
a particular line-of-sight $x$ the weight function ${\cal K}(x,r)$ is
responsible for the fact that not all parts of that line-of-sight
contribute in the same proportion to the dusty projection, as is the
case in the optically thin regime. From equation~(\ref{kingweight})
and appendix~{\ref{appendixA}} we can see that the weight function is
a monotonically rising function of $r$ for all lines-of-sight $x$,
essentially because along a line-of-sight the near side dominates the
light. Dust makes the contribution of the outer parts (in particular
the near side) thus more important, relative to the inner parts.

In figure~{\ref{deptausig.ps}{a,b}} we plot the projected velocity
dispersion profile $\sigma_p^2(x)$ for various optical depths. The
explicit dependence of $\sigma_p^2(x)$ on $\tau$ for a fixed
line-of-sight $x$ is shown in figure~{\ref{deptausig.ps}{c}}. From
these figures it is clear that quantitatively, the effects of dust
absorption on the projected velocity dispersion are rather
limited. Particularly for modest optical depths, $\tau\la2$, there is
no major difference between normal and dusty projection of the
dispersion. In figure~{\ref{deptausig.ps}{d}} we plot the strength (in
percentage)
\begin{equation}
	\Delta_{\sigma}(x)
	\equiv
	100\times\frac{\left.\sigma_p^2(x)\right|_{\tau=0} - 
	\left.\sigma_p^2(x)\right|_{\tau}}
	{\left.\sigma_p^2(x)\right|_{\tau=0}} 
\end{equation}
of the effect as a function of the anisotropy parameter $q$. It is
shown for a set of lines-of-sight, with $\tau=2$. The effect is only
1.6 per cent for the central line-of-sight, independent of the
anisotropy of the model. At larger radii this anisotropy does become
important, tangentially anisotropic systems are more sensitive to dust
in their $\sigma_p^2(x)$ profile than their radial counterparts.  At
$x=r_c$ this drops to 1.2 per cent for the tangential and to 0.7 per
cent for the radial model.

Qualitatively, the general effect of dust is to lower the projected
dispersion profile, and mostly $\Delta_{\sigma}$ will be a positive
quantity. Only for the outer lines-of-sight of very radially
anisotropic models the effect has the opposite sign, i.e.\ the
projected dispersion is higher for the optically thick than for the
optically thin case. In figure~{\ref{deptausig.ps}{e}} we plot the
critical line-of-sight $x_c$ where $\Delta_{\sigma}(x_c) = 0$ as a
function of $q$, for various values of $\tau$. Notice that its
dependence upon the optical depth is very weak.

\subsubsection{The line profiles}

\begin{figure*}
\centering
\includegraphics[clip,bb=60 271 536 685]{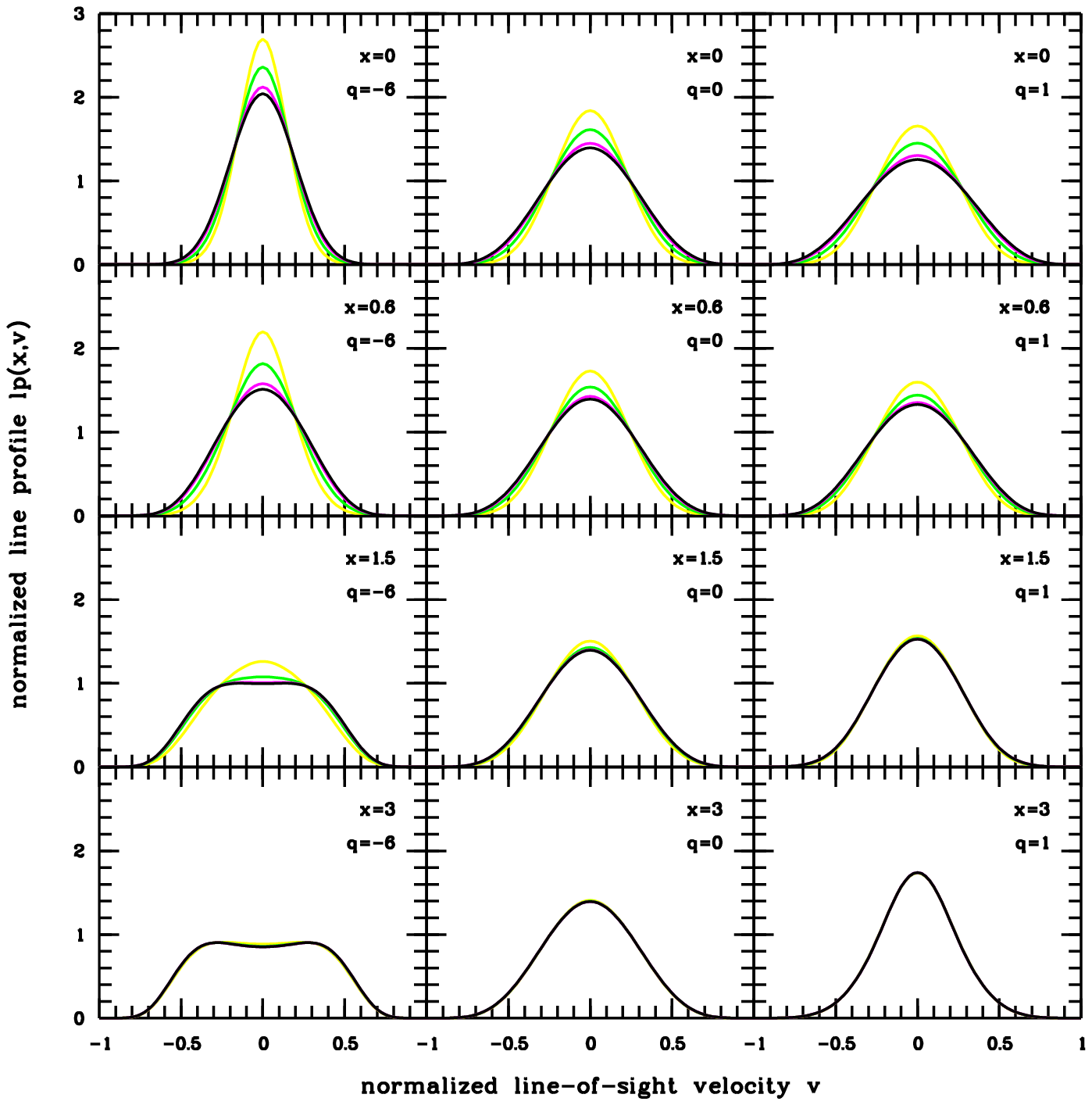}
\caption{A set of normalized line profiles for our standard dust model
($r_0=1$ and $\alpha=3$) for varying optical depth. The left, middle
and right column correspond to respectively the radial, isotropic and
tangential models. The line-of-sight $x$ is indicated. Line profiles
are shown for $\tau=0$, $\tau=4$, $\tau=10$ and $\tau=20$, going from
black to light-coloured grey. Both the line-of-sight velocities and the
area under the curves are normalized, as described in the text. }
\label{deptaulpr.ps}
\end{figure*}

In figure~{\ref{deptaulpr.ps}} a number of normalized line-profiles
${\rm lp}(x,v)$ are plotted, for a set of lines-of-sight varying from
the centre to the outer parts of the galaxy. The line-of-sight
velocities $v$ are normalized with respect to the largest possible
velocity one can observe at the particular line-of-sight $x$, being
the escape velocity $\sqrt{2\,\psi(x)}$, where $\psi(r)$ is the
Plummer potential~(\ref{plummpair1}). The total area under the line
profile, which equals the zeroth moment $\rho_p(x)$, is scaled to one.

The effect of dust absorption on a normalized line-profile ${\rm
lp}(x,v)$ is at first order determined by the effect on its second
moment, i.e.\ the projected dispersion $\sigma^2_p(x)$ at that
projected radius. We can therefore copy most of our findings from the
previous section. Only for higher optical depths and the inner
lines-of-sight the influence of dust absorption is significant. In
most cases dust absorption will turn the line profiles more peaked,
and again tangential models are more sensitive to this than radial
ones.

A particular consequence is that the typical bimodal structure of the
outer line profiles of tangential models can disappear (see the
$x=1.5$ line profile in figure~{\ref{deptaulpr.ps}}). Physically this
can be understood in the following way. In a tangentially anisotropic
model, most of the stars move on rather circular orbits, particularly
in the outer parts. Since the density of the Plummer model follows a
strong $r^{-5}$ at large radii, the line profiles are primarily
governed by the stars which move at orbits with radius $r\approx x$
(in both senses), causing the bimodal structure. When the system is
heavily obscured, the nearest parts of the line-of-sight contribute
relatively more than in the optically thin case. And these nearest
parts contain stars moving on nearly circular orbits with larger
radii, which have velocities that are gradually more perpendicular to
the line-of-sight. Moreover these stars have smaller velocities
anyway, since the circular velocity for the Plummer geometry declines
as $r^{-1/2}$ at large radii. Smaller line-of-sight velocities thus
become predominant, which causes the bimodal structure to disappear.

\subsection{Dependence upon dust geometry}

In this section we vary the dust exponent between $\alpha=1$ and
$\alpha=10$, while the optical depth is held fixed. We choose large
values for $\tau$ ($\tau=10$ and $\tau=20$) so that the geometry
effect is clearly visible.

\subsubsection{The projected density}

\begin{figure*}
\centering 
\includegraphics[clip,bb=60 350 536 685]{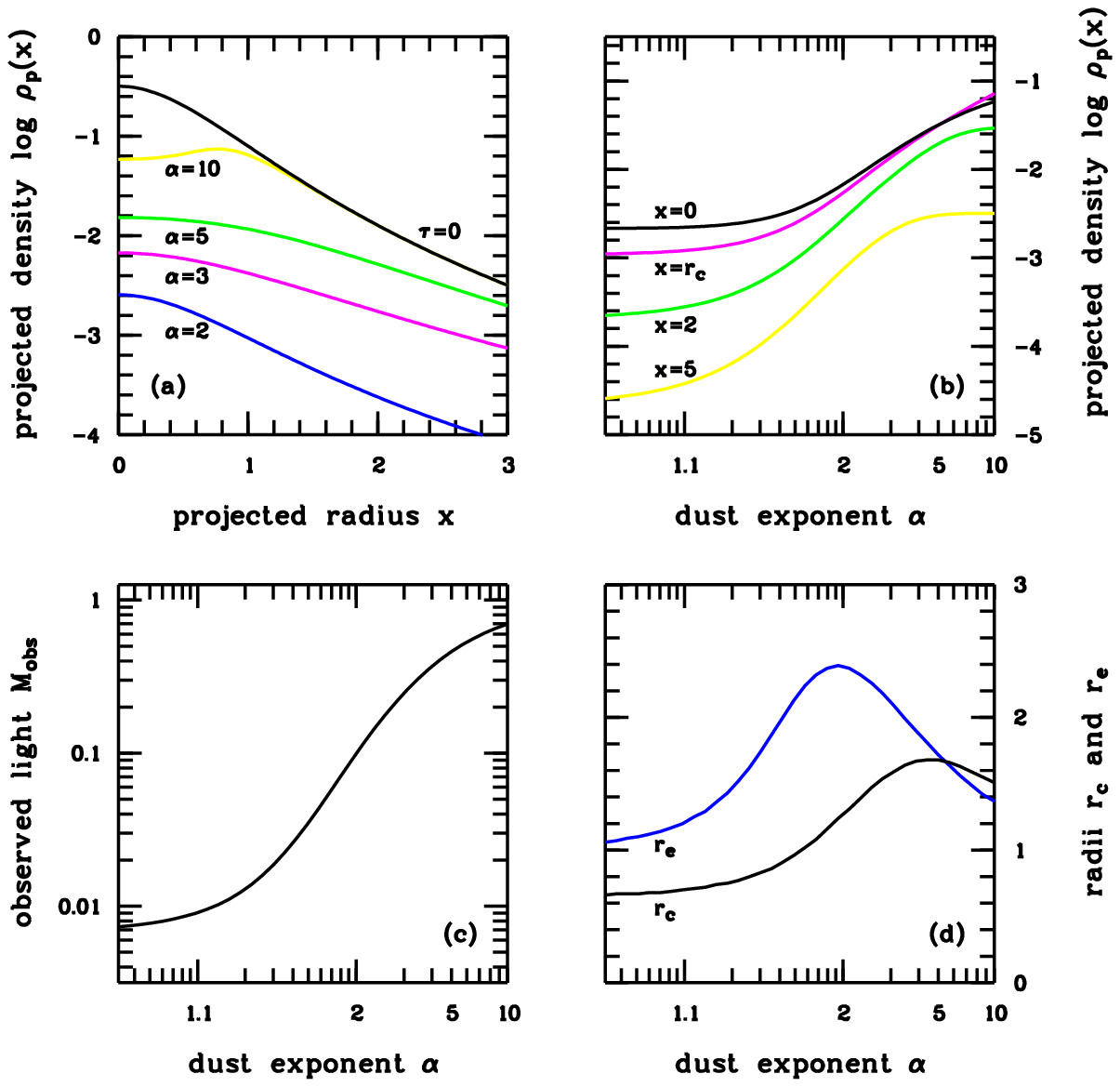}
\caption{The dependence of the projected density $\rho_p(x)$ upon the
dust exponent $\alpha$, with $\tau=10$. In panel~{(a)} the projected
density is plotted as a function of the projected radius $x$ for
various values of $\alpha$, which are indicated. Figure~{(b)} shows
the projected density as a function of $\alpha$, for four
lines-of-sight $x$. The lower two panels show {(c)} the total
integrated light $M_{\rm obs}$, and {(d)} the core radius $r_c$ and
the effective radius $r_e$ as a function of $\alpha$.}
\label{depalprho.ps}
\end{figure*}

The dependence of $\rho_p(x)$ on the dust exponent $\alpha$ is shown
in figure~{\ref{depalprho.ps}{a}} and~{\ref{depalprho.ps}{b}}, where
we adopt the optical depth $\tau=10$. As is to be expected, the
projected density $\rho_p(x)$ is sensitive to variations of the dust
geometry parameters. In the outer regions (the bottom curves in
figure~{\ref{depalprho.ps}{b}}) this is primarily due to the optical
depth profile~(\ref{optdepth}) which depends critically on
$\alpha$. Smaller values of $\alpha$ mean more dust and therefore more
extinction. But also in the inner regions (black curve in
figure~{\ref{depalprho.ps}{b}}) the extinction effect is larger for
extended than for concentrated distributions, even though equal
amounts of dust are present. At each projected radius $x$ and for each
optical depth $\tau$, the projected density $\rho_p(x)$ is a rising
function of $\alpha$, with lowermost value
\begin{equation}
	\lim_{\alpha\rightarrow1}\rho_p(x) 
	=
	\frac{1}{\pi}\,{\rm e}^{-\frac{\tau}{2}}\,
	\frac{1}{\left(1+x^2\right)^2}.
\end{equation}
As a consequence, the total integrated light $M_{\rm obs}$ increases
with $\alpha$ too (figure~{\ref{depalprho.ps}{c}}), with
$\exp\left(-\tfrac{\tau}{2}\right)$ for the overlying screen model
being the minimum value. The more the dust is distributed
homogeneously over the line-of-sight, the more efficient the
extinction process.

Also the morphology of the $\rho_p(x)$ profile is sensitive for
variations of the dust geometry (figure~{\ref{depalprho.ps}{a}}). For
small values of $\alpha$ the effect is small, and in the limit of the
screen geometry the effect is nil~: the projected density decreases
with the same factor for all lines-of-sight. For larger values of
$\alpha$, the profile becomes flatter in the centre. For very
condensed dust distributions the extinction is confined to the central
parts of the galaxy. 

The dependence of the core radius $r_c$ and the effective radius $r_e$
on $\alpha$ is illustrated in figure~{\ref{depalprho.ps}{d}}. The
maximum effect on these core size parameters is found at intermediate
values of $\alpha$. For very extended distributions on the one hand,
the entire $\rho_p(x)$ profile is affected with the same factor, and
as a consequence the core size parameters are not affected. For very
concentrated distributions on the other hand only the inner part of
the core is affected, so that the total light output of the core
remains nearly the same.

\subsubsection{The projected dispersion}

\begin{figure*}
\centering 
\includegraphics[clip,bb=60 350 536 685]{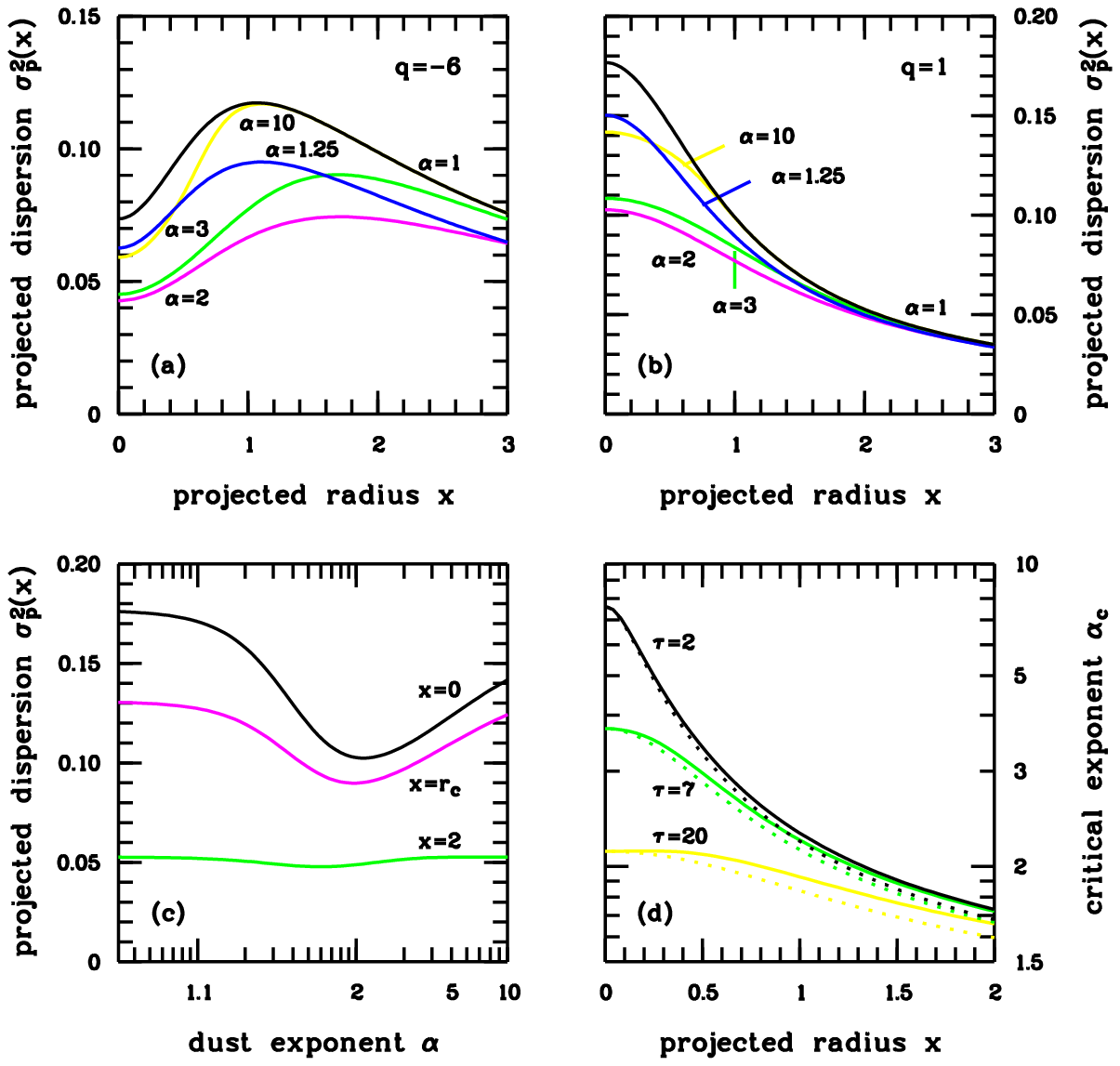}
\caption{The dependence of the projected dispersion $\sigma_p^2(x)$
upon the dust exponent $\alpha$. The optical depth is fixed at the
value $\tau=20$ for these plots. In the upper two panels
$\sigma_p^2(x)$ is plotted as a function of the projected radius $x$
for various dust geometries, which are indicated. It is shown {(a)}
for a tangential model with $q=-6$ and {(b)} a radial model with
$q=1$. In panel~{(c)} we show the projected dispersion explicitly as a
function of the optical depth $\alpha$, for three lines-of-sight
$x$. It is plotted for the radial model $q=1$. In plot~{(d)} we plot
the critical dust exponent $\alpha_c$ (see text) as a function of the
projected radius $x$ for various optical depths and orbital
models. The values of $\tau$ are indicated. The solid lines correspond
to the radial $q=1$ model, the dotted lines to the tangential $q=-6$
models.  }
\label{depalpsig.ps}
\end{figure*}

The primary effect of dust absorption on the projected dispersion
$\sigma^2_p(x)$ at a certain line-of-sight $x$ is to attach a weight
to every part of that line-of-sight. Or, since both halves of our
Plummer galaxy are dynamically equivalent, to every part of the nearer
half of the line-of-sight. Since the parameter $\alpha$ controls
the relative distribution of the dust along the lines-of-sight, this
parameter determines in the first place the relative weights of the
different parts.

There are two extreme cases. For very concentrated models on the one
hand, the entire first half of the line-of-sight is not affected. For
very extended models on the other hand, the dust is effectively
located between the observer and the galaxy, and all the parts of the
line-of-sight $x$ experience the same influence, and thus have the
same relative weight. We obtain that for both cases, each part of the
nearer half of the line-of-sight contributes equally to the projected
dispersion, as in the optically thin limit. The dusty projected
dispersion profile $\sigma^2_p(x)$ therefore equals the optically thin
profile~(\ref{plummer:projdisp}), no matter how large the optical
depth $\tau$.

For the dust models considered here, the relative weights do differ
from each other, and there is an effect on the projected dispersion
profile. In the figures~{\ref{depalpsig.ps}a}
and~{\ref{depalpsig.ps}b} we plot the $\sigma_p^2(x)$ profile for
various values of $\alpha$, for a fixed optical depth $\tau=20$. For
every line-of-sight there is a critical exponent $\alpha_c$ that
corresponds to the most efficient dust geometry. This is illustrated
in figure~{\ref{depalpsig.ps}{c}}, where $\sigma_p^2(x)$ is plotted as
a function of $\alpha$ for various lines-of-sight. Notice that the
position of the critical dust exponent $\alpha_c$ not only depends on
the line-of-sight $x$, but also on the optical depth $\tau$ and the
anisotropy parameter $q$. This is shown in
figure~{\ref{depalpsig.ps}{d}}, where $\alpha_c$ is plotted as a
function of $x$ for different values of $\tau$ and~$q$.

\subsubsection{The line profiles}

\begin{figure}
\centering
\includegraphics[clip,bb=181 525 414 686]{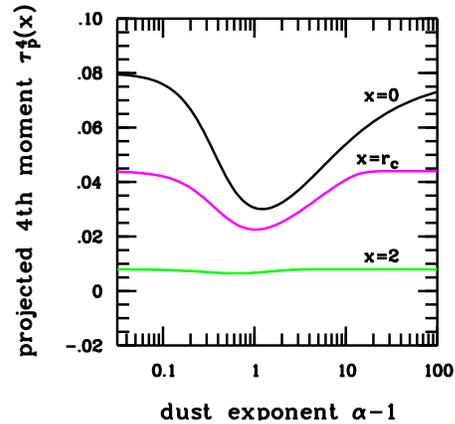}
\caption{The projected fourth moment $\tau_p^4(x)$ as a function of
the dust exponent $\alpha$ for a set of lines-of-sight $x$. The
optical depth is $\tau=20$, the model is the radial $q=1$
model.}
\label{depalptau.ps}
\end{figure}

\begin{figure*}
\centering
\includegraphics[clip,bb=60 178 536 685]{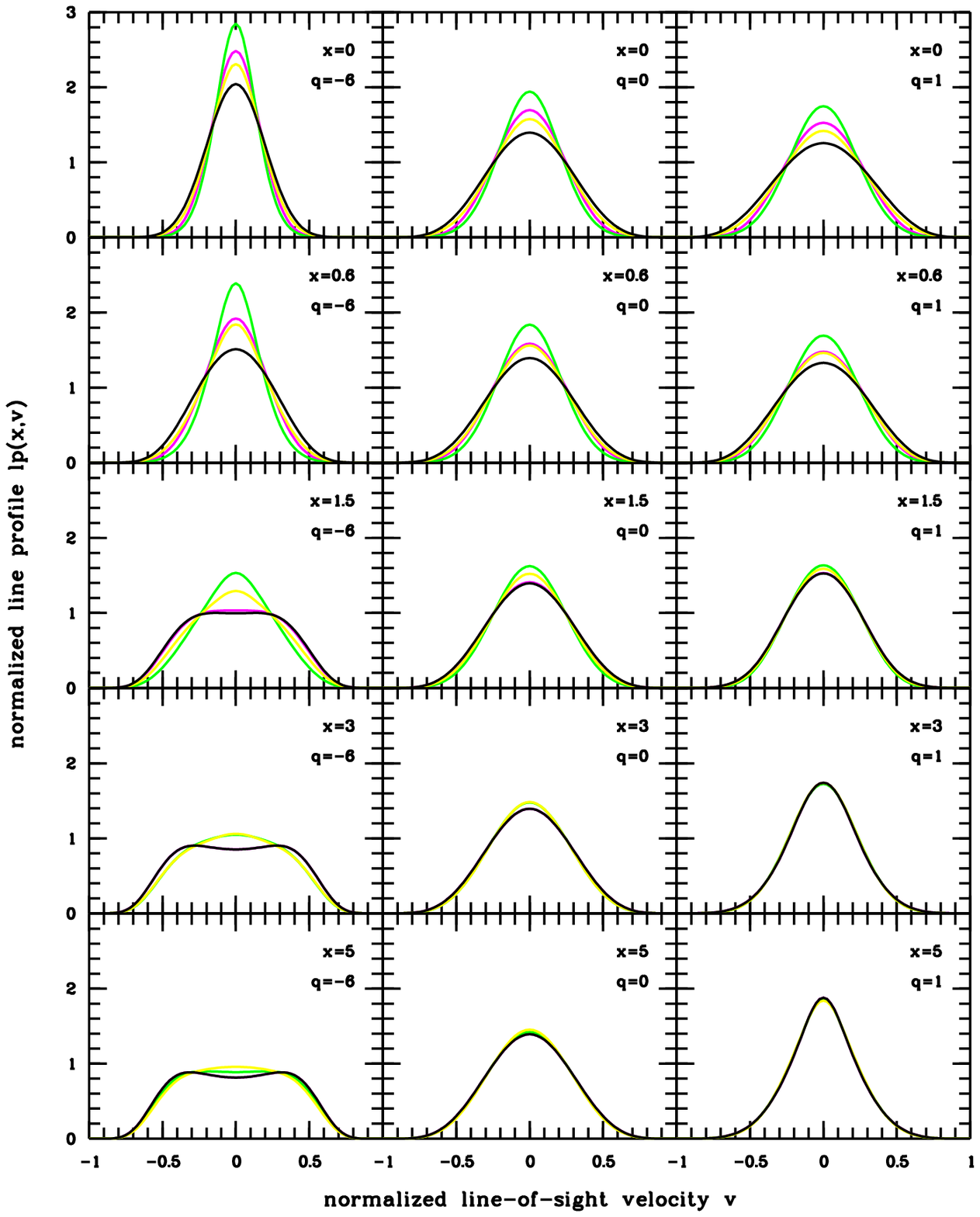}
\caption{A set of normalized line profiles ${\rm lp}(x,v)$ for various
values of the dust exponent $\alpha$ and a fixed optical depth
$\tau=20$. The left, middle and right column correspond to
respectively the radial, isotropic and tangential models. The
line-of-sight $x$ is indicated. Line profiles are shown for the
optically thin model and for $\alpha=5$, $\alpha=2$ and $\alpha=1.25$,
going from black to light-coloured grey. Both the line-of-sight
velocities and the area under the curves are normalized. }
\label{depalphalpr.ps}
\end{figure*}

The reasoning from the previous section does not only apply to the
projected dispersion profile, but to all higher order moments of a
line profile (for a definition, see appendix~{\ref{appendixC}}). For
each line-of-sight $x$, such a higher order projected moment equals
its optically thin value for very concentrated or very extended
models, and there is an intermediate critical value $\alpha_c$ for
which the effect is maximal. This is illustrated in
figure~{\ref{depalptau.ps}}, where we plot the projected fourth moment
$\tau_p^4(x) \equiv \mu_p^4(x) / \rho_p(x)$ for various
lines-of-sight, as a function of the dust exponent $\alpha$. The
analogy with figure~{\ref{depalpsig.ps}{c} is obvious.  Notice that
the critical dust exponent $\alpha_c$ differs from moment to
moment. For example, for the line-of-sight through the centre we find
$\alpha_c=2.115$ for the projected dispersion, and $\alpha_c=2.215$
for the fourth moment.

The same applies to the line profiles themselves
(figure~{\ref{depalphalpr.ps}}). For the two limit cases, there is no
difference between optically thick and optically thin models, no
matter how high the optical depth. And for each line-of-sight $x$,
each optical depth $\tau$ and each orbital model $q$ there is a
critical dust exponent $\alpha_c$, for which the effect of dust
absorption on the line profiles is maximal. The extent to which
$\alpha$ differs from this critical value determines the magnitude of
the effect on the line profile.

\section{Conclusions}

Comparison of optical and {\em IRAS} data suggests that elliptical
galaxies are likely to contain a diffuse dust component, which is
smoothly distributed over the galaxy (Goudfrooij \& de Jong
1995). Dust absorption affects the projection of all physical
quantities along a line-of-sight. We constructed a four-parameter
model to investigate the absorption effects on the projection of
kinematic quantities. We compared the optically thick with the
optically thin case, and investigated the effects of optical depth and
dust geometry.

We established that the projected mass density $\rho_p(x)$ is
substantially affected, even for modest optical depths
($\tau\la2$). Dust has also a non-negligible effect on the morphology
of the $\rho_p(x)$ profile and on all global photometric
quantities. On the other hand, the effects of dust absorption on the
projected velocity dispersion profile $\sigma^2_p(x)$ and on the line
profiles are considerable only for higher optical depths
($\tau\ga5$). Moreover, the dust geometry is important~: both
centrally concentrated and very extended dust distributions can remain
invisible in the line profiles.

Our models have some important applications, which are being
investigated~:
\begin{itemize}
\item Due to the insensitivity of the $\sigma_p^2(x)$ profile,
dynamical mass estimates will hardly be affected by modest amounts of
dust, while the light profile is sensitive for absorption. This should
be taken in consideration in the interpretation of mass-to-light
ratios in regions where no dark matter is expected~: large
mass-to-light ratios can indicate dust absorption.
\item All global photometric properties of elliptical galaxies such as
luminosities and radii are affected by dust. As WS96 pointed out, this
can lead to a systematic uncertainty in elliptical galaxy distance
estimation techniques that utilize these global properties, e.g.\ the
$L$-$\sigma$ relation (Faber \& Jackson 1976) or the relations of the
fundamental plane (Dressler {\em et al.\ }1987, Djorgovski \& Davis
1987). With our models, we can not only quantify the effect of dust on
the global photometric properties, but also on the kinematics.
\item While large optical depths don't seem to be in accordance with
global {\em IRAS} dust mass estimates, these models are nevertheless
important on a local scale. Compact regions with high optical depths
can be an explanation for the observed asymmetries and irregularities
in kinematic data. In that case, one needs to combine (basically
dust-free) near-infrared data with optical data. Such observations
would help to constrain the diffuse dust masses in elliptical
galaxies, a problem which cannot be satisfyingly solved with the
existing methods.
\end{itemize}

\appendix

\section{The function ${\bf U}_\alpha(\lowercase{z})$}
\label{appendixA}

\begin{figure}
\centering
\includegraphics[clip,bb=179 488 416 681]{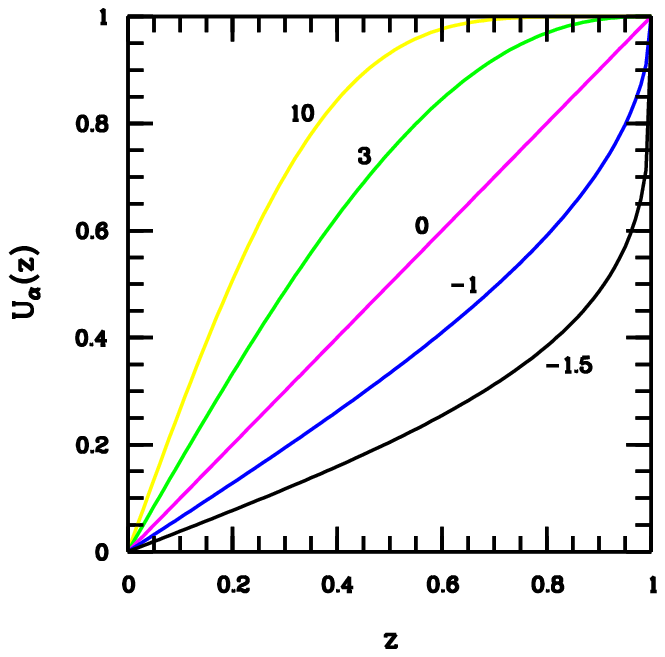}
\caption{The function ${\bf U}_\alpha(z)$ for different values of
$\alpha$. The values of $\alpha$ are indicated.}
\label{ufunc.ps}
\end{figure}

For $\alpha>-2$ and for $0\leq z \leq1$ the function ${\bf
U}_\alpha(z)$ is defined as
\begin{align}
	{\bf U}_\alpha(z) 
	&= 	
	\int_0^z\left(1-y^2\right)^{\frac{\alpha}{2}}\,{\rm d}y
	\left/
	\int_0^1\left(1-y^2\right)^{\frac{\alpha}{2}}\,{\rm d}y.
	\right.
	\nonumber \\
	&= 	
	\frac{2}{\sqrt{\pi}}\, 
	\frac{\Gamma\left(\frac{\alpha+3}{2}\right)}
	     {\Gamma\left(\frac{\alpha+2}{2}\right)}\,
	\int_0^z\left(1-y^2\right)^{\frac{\alpha}{2}}\,{\rm d}y
	\label{U_definition}
\end{align}
For natural values of $\alpha$ it can be easily calculated with the
recursion formula
\begin{equation}
	{\bf U}_\alpha(z) 
	=
	{\bf U}_{\alpha-2}(z) +
	\frac{1}{\sqrt{\pi}}\, 
	\frac{\Gamma\left(\frac{\alpha+1}{2}\right)}
	     {\Gamma\left(\frac{\alpha+2}{2}\right)}\,
	z\,\left(1-z^2\right)^{\frac{\alpha}{2}}
\end{equation}
together with the seeds
\begin{gather}
	{\bf U}_{-1}(z) = \frac{2}{\pi}\,\arcsin(z)
	\\
	{\bf U}_0(z) = z.
\end{gather}
The functions ${\bf U}_\alpha(z)$ can also be written in terms of
the hypergeometric function, which is convenient for computational
purposes if $\alpha$ is non-integer,
\begin{equation}
	{\bf U}_\alpha(z)
	=
	\frac{2}{\sqrt{\pi}}\, 
	\frac{\Gamma\left(\frac{\alpha+3}{2}\right)}
	     {\Gamma\left(\frac{\alpha+2}{2}\right)}\,
	z\,_2F_1\left(
	\tfrac{1}{2},
	-\tfrac{\alpha}{2};
	\tfrac{3}{2};
	z^2\right). 
\end{equation}
In the limit $\alpha\rightarrow-2$ one finds the degenerate function
\begin{equation}
	\lim_{\alpha\rightarrow-2}{\bf U}_\alpha(z)
	=
	\left\{\begin{array}{l@{\qquad}l}
	0 & \mbox{for $0\leq z<1$} \\
	1 & \mbox{for $z=1$.}
	\end{array}\right.
\end{equation}

\section{The moments of the central line profile}
\label{appendixC}

\begin{table*}
\caption{Analytical expressions for the central projected density
$\rho_p(0)$ and the central projected dispersion $\sigma_p^2(0)$ as a
function of the optical depth $\tau$. In all cases $r_0=1$, the value
of $\alpha$ is indicated. The expressions for $\alpha=3$ and
$\alpha=2$ correspond to equations~(\ref{momalp3})
and~(\ref{momalp2}). For $\alpha=5$ the projected density can be
calculated by direct integration, for higher order moments no explicit
expression can be deduced.}
\label{central.tbl}
\centering
\begin{tabular}{@{}c@{\qquad\quad}c@{\qquad\quad}c@{}}
\hline\hline
$\alpha$ & $\rho_p(0)$ & $\sigma_p^2(0)$\\
\hline
1 & $\displaystyle \frac{1}{\pi}\,{\rm e}^{-\frac{\tau}{2}} $ &
	$\displaystyle \left(\frac{1}{6-q}\right)\,
	\frac{9\pi}{32}$ \\ \\	
2 & $\displaystyle 9\pi^3\,{\rm e}^{-\frac{\tau}{2}}\,
	\frac{\cosh\left(\frac{\tau}{2}\right)}
	{(\tau^2+\pi^2)\,(\tau^2+9\pi^2)}$ &
	$\displaystyle \left(\frac{1}{6-q}\right)\,\pi\,
	\tanh\left(\tfrac{\tau}{2}\right)\,
	\frac{(\tau^2+\pi^2)\,(\tau^2+9\pi^2)}
	{\tau\,(\tau^2+4\pi^2)\,(\tau^2+16\pi^2)}$ \\ \\
3 & $\displaystyle \frac{6}{\sqrt{\pi}}\,{\rm e}^{-\frac{\tau}{2}}\,
	\tau^{-\frac{3}{2}}\,
	I_\frac{3}{2}\!\left(\tfrac{\tau}{2}\right)$ &
	$\displaystyle \left(\frac{1}{6-q}\right)\,
	\frac{3\sqrt{\pi}}{2}\,\frac{1}{\sqrt{\tau}}\,
	\frac{I_2\!\left(\frac{\tau}{2}\right)}
	{I_\frac{3}{2}\!\left(\frac{\tau}{2}\right)}$ \\ \\
5 & $\displaystyle \frac{1}{\pi}\,\frac{1-{\rm e}^{-\tau}}{\tau}$ & \\ \\
\hline
\end{tabular}
\end{table*}

The general moments $\mu_{2n,2m,2l}(r)$ of a distribution function
$F(r,\bfv)$ are defined as
\begin{equation}
	\mu_{2n,2m,2l}(r)
	=
	\int\!\!\int\!\!\int F(r,\bfv)\,
	v_r^{2n} v_\theta^{2m} v_\phi^{2l}
	\,{\rm d}\bfv,
\end{equation}
where the integration ranges over all possible velocities. The moments
$\mu_{2n}(r,x)$ of the projected velocity distribution along a
line-of-sight $x$ at a particular point $(x,r)$ in space are
\begin{equation}
	\mu_{2n}(x,r)
	=
	\sum_{i=0}^{n}
	\binom{n}{i}\,
	\frac{\left(r^2-x^2\right)^{n-i}\,x^{2i}}{r^{2n}}\,
	\mu_{2(n-i),2i,0}(r).
\label{musum}
\end{equation}
To obtain the $2n$'th moment of the line profile (or the projected
$2n$'th moment) this has to be projected along the line-of-sight $x$,
according to the recipe~(\ref{dusty_projection},b),
\begin{equation}
	\mu_p^{2n}(x)
	\equiv
	2\,\int_x^{+\infty}{\cal K}(x,r)\,
	\frac{\mu_{2n}(x,r)\,r\,{\rm d}r}{\sqrt{r^2-x^2}}.
\label{momp}
\end{equation}
For the Plummer model, these projected moments can be evaluated
analytically in the optically thin limit (Dejonghe 1987),
\begin{multline}
	\mu_p^{2n}(x)
	=
	\frac{3\cdot2^{n-2}}{\pi}\,
	\frac{\Gamma\left(6-q\right)\,\Gamma\left(n+\frac{1}{2}\right)\,
	\Gamma\left(\frac{n+4}{2}\right)}
	{\Gamma\left(n+6-q\right)\,\Gamma\left(\frac{n+5}{2}\right)}
	\left(1+x^2\right)^{-\frac{n+4}{2}}
	\\ \times
	\!\,_3F_2\left(
	-n,\frac{q}{2},\frac{n+4}{2}; 1,\frac{n+5}{2},
	\frac{x^2}{1+x^2}\right).
\label{mupalg}
\end{multline}	
yielding equation~(\ref{plummer:projdens})
and~(\ref{plummer:projdisp}) for the cases $n=0$ and $n=1$.

For dusty galaxy models, equation~(\ref{momp}) cannot be calculated
analytically for general lines-of-sight.\footnote[1]{Except for the
degenerate case $\alpha=1$, where $\mu_p^{2n}(x)$ equals the product
of the optically thin moment~(\ref{mupalg}) and the weight
function~(\ref{screenweight}).} Only for the central line-of-sight
$x=0$ this is possible in a few cases, since the formulae become
considerably easier. The sum~(\ref{musum}) reduces to a single term,
\begin{align}
	\mu_{2n}(0,r) 
	&= 
	\mu_{2n,0,0}(r) \\
	&=	
	\frac{3\cdot2^{n-2}}{\pi\sqrt{\pi}}\,
	\frac{\Gamma\left(n+\frac{1}{2}\right)\,
	\Gamma\left(6-q\right)}{\Gamma\left(6-q+n\right)}\,
	\left(1+r^2\right)^{-\frac{n+5}{2}},
\end{align}
yielding equation~(\ref{plummpair2}) and~(\ref{raddisp}) for $n=0$ and
$n=1$ respectively. If we substitute the weight
function~(\ref{kingweight}) of the King dust model, we obtain
\begin{equation}
	\mu_p^{2n}(0)
	=
	c\,{\rm e}^{-\frac{\tau}{2}}\,
	\int_0^1
	\cosh\left(\tfrac{\tau}{2}\,
	{\bf U}_{\alpha-3}(z)\right)
	\left(1-z^2\right)^{\frac{n+2}{2}}{\rm d}z
\label{wawa}
\end{equation}
where we set
\begin{equation}
	c = 
	\frac{3\cdot2^{n-1}}{\pi\sqrt{\pi}}\,
	\frac{\Gamma\left(6-q\right)\,\Gamma\left(n+\frac{1}{2}\right)}
	{\Gamma\left(n+6-q\right)}.
\end{equation}
Since ${\bf U}_{\alpha-3}(z)$ can generally only be expressed as a
hypergeometric function, we cannot evaluate this integral explicitly,
except for some special values of the dust exponent $\alpha$ for which
${\bf U}_{\alpha-3}(z)$ has a very simple form. For $\alpha=3$ the
integral in~(\ref{wawa}) leads to a family of modified Bessel
functions of the first kind (Gradshteyn \& Ryzhik 1965, eq.\ 8.431),
\begin{equation}
	\mu_p^{2n}(0)
	=
	4\sqrt{\pi}\,c\,
	\Gamma\left(\tfrac{n+4}{2}\right)\,
	{\rm e}^{-\frac{\tau}{2}}\,\tau^{-\frac{n+3}{2}}\,
	I_{\frac{n+3}{2}}\left(\tfrac{\tau}{2}\right).
\label{momalp3}
\end{equation}
For $\alpha=2$ we find (Gradshteyn \& Ryzhik 1965, eq.\ 3.981)
\begin{equation}
	\mu_p^{2n}(0)
	=
	\left\{
	\begin{array}{l@{\qquad}l}
	\displaystyle
	\frac{\pi^{n+4}\,c\,\Gamma(n+4)\,{\rm e}^{-\frac{\tau}{2}}\,
	\cosh\left(\tfrac{\tau}{2}\right)}
	{\prod_{j=0}^{(n+2)/2} \Bigl(\tau^2+(2j+1)^2\pi^2\Bigr)}
	& \mbox{if $n$ even} \\ \\
	\displaystyle
	\frac{\pi^{n+4}\,c\,\Gamma(n+4)\,{\rm e}^{-\frac{\tau}{2}}\,
	\sinh\left(\tfrac{\tau}{2}\right)}
	{\tau\,\prod_{j=1}^{(n+3)/2} \Bigl(\tau^2+(2j)^2\pi^2\Bigr)}
	& \mbox{if $n$ odd}.
	\end{array}
	\right.
\label{momalp2}
\end{equation}
From these expressions one can find the expressions for the projected
density $\rho_p(0)$ and the projected dispersion $\sigma_p^2(0)$, as
tabulated in table~{\ref{central.tbl}}.

\bsp

\end{document}